# Oxidative Species-Induced Excitonic Transport in Tubulin Aromatic Networks: Potential Implications for Neurodegenerative Disease

P. Kurian[1,2,3], T. O. Obisesan[4], T. J. A. Craddock[5,6]

[1]National Human Genome Center, Howard University College of Medicine, Washington, DC 20060, USA
[2]Department of Medicine, Howard University College of Medicine, Washington, DC 20060, USA
[3]Computational Physics Laboratory, Howard University, Washington, DC 20059, USA
[4]Georgetown-Howard Universities Center for Clinical and Translational Science Clinical Research Unit, Howard University College of Medicine, Washington, DC 20060, USA
[5]Departments of Psychology & Neuroscience, Computer Science, and Clinical Immunology, Nova Southeastern University, Fort Lauderdale, FL 33314, USA
[6]Clinical Systems Biology Group, Institute for Neuro-Immune Medicine, Nova Southeastern University, Fort Lauderdale, FL 33328, USA

**ABSTRACT**

Oxidative stress is a pathological hallmark of neurodegenerative tauopathic disorders such as Alzheimer's disease and Parkinson's disease-related dementia, which are characterized by altered forms of the microtubule-associated protein (MAP) tau. MAP tau is a key protein in stabilizing the microtubule architecture that regulates neuron morphology and synaptic strength. When MAP tau is degraded in tauopathic disorders, neuron dysfunction results. The precise role of reactive oxygen species (ROS) in the tauopathic disease process, however, is poorly understood. Classically, mitochondrial dysfunction has been viewed as the major source of oxidative stress and has been shown to precede tau and amyloid pathology in various dementias, but the exact mechanisms are not clear. It is known that the production of ROS by mitochondria can result in ultraweak photon emission (UPE) within cells. While of low intensity, surrounding proteins within the cytosol can still absorb these energetic photons via aromatic amino acids (e.g., tryptophan and tyrosine). One likely absorber of these photons is the microtubule cytoskeleton, as it forms a vast network spanning neurons, is highly co-localized with mitochondria, and shows a high density of aromatic amino acids. Functional microtubule networks may traffic this ROS-generated endogenous photon energy for cellular signaling, or they may serve as dissipaters/conduits of such energy to protect the cell from potentially harmful effects. Experimentally, after *in vitro* exposure to exogenous photons, microtubules have been shown to reorient and reorganize in a dose-dependent manner with the greatest effect being observed around 280 nm, in the tryptophan and tyrosine absorption range. In this paper, recent modeling efforts based on ambient temperature experiment are presented, showing that tubulin polymers can feasibly absorb and channel these photoexcitations via resonance energy transfer, on the order of dendritic length scales and neuronal fine structure. Since microtubule networks are compromised in tauopathic diseases such as Alzheimer's and Parkinson's dementias, patients with these illnesses would be unable to support effective channeling of these photons for signaling or dissipation. Consequent emission surplus due to increased UPE production or decreased ability to absorb and transfer may lead to increased cellular oxidative damage, thus hastening the neurodegenerative process.

# INTRODUCTION

There exists a group of heterogeneous neurodegenerative disorders that are characterized by significant intracellular accumulations of neurofibrillary tangles (NFTs), abnormal filaments formed by the intrinsically disordered microtubule-associated protein (MAP) tau. Collectively, these diseases are labeled as neurodegenerative tauopathies and include Alzheimer's disease (AD), fronto-temporal lobar degeneration with tau inclusions, progressive supranuclear palsy, corticobasal degeneration, Parkinson's disease-related dementia, and chronic traumatic encephalopathy (CTE), among other dementias [1]. MAP tau stabilizes the microtubule cytoskeleton in neuronal axons, and when it is compromised the microtubule cytoskeleton disintegrates, leading to loss of neuron structure and neurodegeneration (see Figure 1). While the stability of neuronal microtubules is a target for prevention and treatment of these disorders [1-4], understanding the mechanisms by which the microtubule cytoskeleton may be disturbed can lead to novel diagnostics and therapies for tauopathic illness.

A prominent early event in the pathogenesis of tauopathy is oxidative stress, which contributes to tau phosphorylation and the formation of NFTs [5]. However, the relationship between intracellular reactive oxygen species (ROS) generated by oxidative stress and tau hyperphosphorylation remains unclear. Accumulation of hyperphosphorylated tau has been shown to cause oxidative stress, while simultaneously ROS have been shown to stimulate tau hyperphosphorylation [6] in a reinforcing positive feedback manner. Surplus or deficient levels of ROS can drive the catastrophic destabilization or abnormal polymerization of microtubules [7-8], which may in turn stimulate the hyperphosphorylation of tau. As pathways leading from soluble, monomeric to insoluble, hyperphosphorylated tau protein are at the center of many human tauopathies, mechanism-based therapies often focus on the prevention of tau aggregation and propagation [9]. However, modulating oxidative stress is another potential pathway towards suppression of tau accumulation and accompanying neurotoxicity [10-11]. Yet there are still other mechanisms by which ROS can interact with proteins in general and microtubules specifically.

Aerobic life continually produces ROS, which can oxidize biomolecules and produce alterations in DNA, proteins, and lipids, particularly in the pathogenesis of age-related disease. ROS free radicals can interact with protein structures leading to the oxidation of thiol groups or glutathionylation, thus modulating their activity. However, one byproduct of ROS reactions includes the generation of singlet oxygen and triplet carbonyl groups [12-13]. When these excited-state species relax, they release photons of specific energies. These photon signals are mostly observed in the visible spectrum, often showing a broad peak in the 500-700 nm region, which overlaps with the emission wavelength of triplet carbonyl and the dimol reaction of singlet oxygen. Historically controversial works of Gurwitsch suggest that there exist photon emissions from cells lying in the range of the ultraviolet (UV) wavelength of light (190 - 250 nm) [14], with some modern evidence for longer UV wavelengths (250 - 390 nm) [15-19] in biological photon emission. Other researchers have reported measurements of photons throughout the UV, visible, and infrared (IR) parts of the electromagnetic spectrum [20-22] and in tumor cells as energetic as 210 nm [23].

Furthermore, microtubules have been shown to reorient and reorganize in a dose-dependent manner after exposure to UV light, with the greatest effect being observed around 280 nm [24-26]. This is expected to be mediated by the aromatic amino acids tryptophan, tyrosine, and to a lesser degree phenylalanine, as they contribute most strongly to protein absorption in this range. Because carbonyl groups are excited in this same range (e.g., ketones and aldehydes show a weak n → π* absorption at ~270-300 nm) by ROS, they may transfer energy to aromatic amino acid networks in tubulin subcomponents of neuronal microtubules through a non-radiative mechanism (see Figure 2).

While absorption, emission, and fluorescence of biophotons, including UPEs, has a long history of study (see Discussion), the question of whether coherent energy transfer in tubulin and microtubules has a biological role remains open. Rapid signaling through aromatic amino acid networks (e.g., tryptophan, tyrosine) may coordinate the complex organization of the microtubule cytoskeleton, which is required for the tasks of cell division, motor protein trafficking, and motility. Such a signaling mechanism may also explain the observed, apparently UV-mediated, cell-to-cell influence on cell division [27].

An initially excited donor chromophore can convey its electronic energy to an acceptor chromophore via electrodynamic coupling of their transition electric dipole moments owing to the close correspondence, or "resonance," between their energy levels. This resonance energy transfer (RET), which was originally pioneered by Förster (and termed FRET), remains the dominant theory applied in electronic energy transport. However, this process is not unique to photosynthesis. Collections of dye molecules known as J-aggregates are an artificial form of light-harvesting complex (LHC) capable of capturing and manipulating photon energy. At its core, this phenomenon is attributable to the optimal packing of chromophores with a significant transition dipole [28].

Based on initial analyses [29], the unique tryptophan network within an individual tubulin dimer can possess significant transition dipole couplings that are capable of supporting quantum coherent beating effects similar to those observed in the Fenna-Matthews-Olson (FMO) photosynthetic complex [30] and LHCs [31-32]. Furthermore, preliminary results have suggested that this network may support coherent energy transfer at physiological temperature across clusters of tryptophans in tubulin polymers and, by extension, microtubule structures. These aromatic network resonances may be induced by photoexcitation from oxygen byproducts in mitochondria, suggesting a complementary quantum pathway in aerobic respiration to the one that has already been demonstrated in the chemically inverse photosynthetic process.

Functional microtubule networks may use the energy from endogenous UV light as a signaling mechanism throughout the cytoskeleton (see Figure 2). Using a theoretical formalism similar to that used for photosynthetic complexes to investigate energy transfer between tryptophan amino acids in tubulin, Craddock et al. have shown that certain conditions favor this mechanism of signal propagation [29]. This mechanism of energy transfer depends on coupling between the transition dipoles of aromatic amino acid chromophores and theoretically may be disrupted via changes in van der Waals-type London forces. As networks of aromatic amino acids, generating so-called π-π stacking arrangements of delocalized electrons, predominantly stabilize these temporary dispersion forces that induce formation of charge-separated dipoles, excitation of the aromatic network via ROS-induced photoexcitation may serve to alter the stability of microtubules, in normal and pathological conditions.

Persistent electronic coherences may be a general property of any system of compact, nearly static chromophores coupled to the environment. This is exemplified by chlorophyll chromophores in photosynthetic LHCs. The aromatic amino acids tryptophan (Trp) and tyrosine (Tyr) can also provide an ideal source of highly delocalizable π electrons that can interact through the formation of transition-state dipoles upon photoexcitation. Tubulin, the microtubule constituent protein, possesses a network of chromophoric amino acids, which may serve as energy transfer pathways in tubulin and microtubules. The fluorescence quantum yield in water for pure Trp is 0.13 and for pure Tyr is 0.14 [33], and the experimentally observed yield for wild-type tubulin is 0.06 at room temperature. These are comparable to the room-temperature fluorescence quantum yields of bacteriochlorophyll of 0.18, and the yield for LH1 of 0.08 and LH2 of 0.10, respectively.

Here we examine this ROS-driven reaction on microtubule function, as it has bearing on neurodegenerative disease processes. We present simulation efforts for coherent transport of photoexcitation via aromatic amino acids in microtubules in the presence of a noisy environment. Using a combination of homology modeling, molecular dynamics simulations, quantum chemistry, and optical biophysics, we apply structure-based simulations similar to current studies of the FMO complex [34-35] and larger LHCs to probe possible energy transfer mechanisms in tubulin protofilaments, and by extension microtubules. Our results indicate that ROS-induced excitonic propagation via quantum coherent energy transfer in tubulin aromatic networks is at least significant on the micron scale. These simulations could provide quantitatively precise predictions for the role of ROS interaction with the microtubule cytoskeletal architecture and may be used as a pre-diagnostic mechanism for neurodegenerative disease.

**RESULTS**

Following previous work in Ref. [29], we simulated coherent energy transfer along a linear polymer chain of tubulin dimers, treating each tubulin as a network of two-level (ground and excited state) chromophoric systems corresponding to the tryptophans (Trps) and tyrosines (Tyrs) in the protein. Our model was constructed through the use of molecular dynamics (MD) simulations to identify the dominant conformation of tubulin in solution, quantum chemistry calculations to determine the excitation energies of aromatic chromophores in tubulin, and theoretical estimates of the coupling between aromatic transition dipoles. This model was then used to predict spectral characteristics of the tubulin protein, which were compared to experiment to refine model parameters. Following parameter refinement, we used the model to simulate exciton propagation along a linear polymer chain of tubulin dimers.

**Model Construction:** The dominant spatial distribution and orientation of the aromatic Trp residues in tubulin after 17 ns of MD simulation are shown and described fully in Ref. [29]. The addition of Tyr increases the density of chromophore packing in the protein, relative to Trp alone, as reported in Ref. [29]. Nearest-neighbor distances between Trp and Tyr residues in the aromatic network of a single tubulin start from three ångströms (0.3 nm), close enough for London-force dipole coupling (see Figure 3). Importantly, this range includes the distances between the bilin chromophores in cryptophyte marine algae, which has been shown to support quantum-coherent transfer of electronic excitation [28]. This spatial range—i.e., the average separation between neighboring dipoles—is considered much larger than the extent of the aromatic transition densities, so the point-dipole (PD) approximation was used to determine the inter-chromophore couplings. While at the lower end of this range the PD approximation becomes questionable, and it may be more appropriate to use an extended dipole description, here we use the PD approximation as a first-order estimate of the contributions of these effects. As the majority of couplings are much larger than the extent of the aromatic transition densities, the PD approximation is valid for the majority of couplings.

The achievement of a deep theoretical molecular-level understanding of electronic energy transport represents an important challenge in the field, and this includes the treatment of environmental screening effects. The screening and local field effects due to the surrounding protein environment composed of other amino acids were taken into account through modifications to coupling between the aromatic transition dipoles. However, in calculating the excitonic interaction (transition dipole coupling) coefficients, the choice of the dielectric constant value and the method used for screening and local field effects are no trivial matters. Typically, effective protein dielectric constant values are chosen between 2.0 and 4.0; however, experimental measurements of the dielectric constants in protein hydrophobic pockets indicate values well above this range, contravening the conventional practice. The experimentally measured tubulin optical dielectric value of 8.41 [36] also falls well outside this range and has been shown to have a significant impact on coupling strengths [29]. There have been multiple measures and much controversy surrounding this measurement, as recent studies have shown that the dielectric constant can range between 2.56 and 8.41 [37].

The problem of choosing the dielectric constant is further confounded by the expression used to describe the screening and local field effects due to the protein matrix, $f$. The Onsager local field factor is the oft-used local field factor [38, 39], but common methods used in photosynthesis to describe $f$ are the Förster [40] and Juzeliūnas and Andrews [41-42] methods. However, for the range of dielectrics from 2.0 - 8.41 examined in this paper, the Onsager screening term only gives a slight increase in coupling strength over the traditional Förster term, and therefore the results in using these factors are expected to be almost equivalent due to their very close similarity. In the Förster approximation, which accounts only for electrostatic screening effects due to the polarizable medium present between the dipoles, the dipolar couplings are small, indicating little to no interaction between chromophores. This is consistent with previous estimates of Förster dipole – dipole energy transfer between Trps in tubulin [29]. In the simplest approximation due to Förster, $f = 1/\varepsilon_r$ and accounts only for screening effects using the frequency-dependent, macroscopic relative dielectric permittivity $\varepsilon_r$ satisfying the Clausius-Mossotti relation. Juzeliūnas and Andrews [41-42] used a quantum field theoretic approach to derive these screening and local field effects, which provides a more detailed description of the interaction than the simple screening effect described by the Förster method. The Juzeliūnas and Andrews (hereafter Juzeliūnas) approach accounts for screening effects of the medium as well as local field effects yielding $f = \varepsilon_r^{-1}[(\varepsilon_r+2)/3]^2$ in the quantum electrodynamics approximation [41-43].

For the experimentally measured optical dielectric constant ($\varepsilon_r$ = 8.41), coupling strengths between Trps up to 60 cm$^{-1}$ are found, comparable to photosynthetic couplings, with the most significant couplings between Trps αW-346 and αW-388, Trps αW-407 and βW-346, and Trps βW-103 and βW-407. Couplings up to 330 cm$^{-1}$ are found between Trps and Tyrs in the chromophore network. Calculated site energies for all Trp and Tyr chromophores (126) in a three-tubulin polymer range from 35,783 to 40,310 cm$^{-1}$, giving a maximum energy difference of 4,527 cm$^{-1}$. In the central tubulin alone, the Trp site energies range from 36,332 to 36,976 cm$^{-1}$, with a maximum energy difference more than 1.5 times the maximum difference in a bare tubulin dimer [29]. This disparity suggests the alteration of the energy landscape by the polymerization of tubulin.

**Parameter Refinement:** As stated above, the choice of dielectric constant and the method used to account for screening and local field effects are not trivial matters. Thus, for comparison, we consider the different values of 2.0, 4.0, and 8.41 in our refinement of parameters. However, accounting for electrostatic screening effects of the medium as well as local inhomogeneous field effects yields different results depending on the dielectric constant chosen. As the dielectric constant value increases between 2.0 and 8.41, the coupling strengths become increasingly non-negligible. Because the scaling factor $f$ exceeds 1 with $\varepsilon_r$ = 8.41 in the Juzeliūnas approximation, we have also considered the case where all values of $f$ are set equal to unity ("Nofactor"). For further commentary on the choice of $\varepsilon_r$ and $f$, see Ref. [29]. As a measure for comparison, we align the absorption, circular dichroism (CD), and linear dichroism (LD) spectra predicted from our modeling of a single tubulin dimer with experimental values. It is clear from Figure 4 that the general line shape of the theoretical spectra for $\varepsilon_r$ values between 2.0 and 4.0 does not align well with experiment. When we use the measured dielectric value of tubulin ($\varepsilon_r$ = 8.41), the theoretical curves come into much closer alignment with experimental values, particularly in the cases of absorption and LD. This same closer alignment at $\varepsilon_r$ = 8.41 is also observed for the CD spectrum when considering only Trp chromophores (centered at 35088 cm$^{-1}$ or 285 nm), but this agreement is disturbed when Tyr chromophores (centered at 35714 cm$^{-1}$ or 280 nm) are added. Similar issues with CD spectra have been observed by other authors [44] attempting to study multi-chromophore systems, and these issues have been mitigated, but not resolved, using adjustable shift parameters, Gaussian broadening with judiciously chosen width parameters (here we used 3500 cm$^{-1}$ full-width half maximum for all chromophores), and extended transition charge distributions. This last approach may be particularly relevant for the addition of Tyrs to our model, as the increased clustering of aromatics (from 8 to 42 total) within a single tubulin causes the point-dipole approximation to break down for a larger fraction of the inter-chromophore couplings. Pardoning this exception, and based on the close agreement with experiment shown in Figure 4, we have chosen $\varepsilon_r$ = 8.41 with a variety of scaling factors (Förster, Juzeliūnas, Nofactor) and dephasing parameters (50, 150, 300 cm$^{-1}$) in the simulations below.

**Exciton Propagation Simulations:** Similar to other models of environmental noise-assisted transport [45], alternating cycles of coherent delocalization and incoherent "hopping" due to localization and decoherence, on the order of tens to thousands of femtoseconds, may work together to extend the lifetime of excitonic propagation, all before the fluorescence time of tryptophan (~1 ns). Such a dynamic interplay between coherent and incoherent energy transfer may represent the "poised realm" [46] at the boundary between the quantum and classical worlds.

Simulations were run on a linear polymer chain of tubulin by constructing a triplet of tubulins and, due to the translational symmetry along the polymer, making calculations always from the chromophore network of the central dimer. To generate a linear polymer of three tubulin proteins, the tubulin dimer was shifted along the microtubule protofilament axis by multiples of 81.9 ångströms [47-49], using PyMOL v1.8.2.3 [50]. Site energies and couplings were calculated via the same methods as for the single tubulin.

Exciton population dynamics calculations using the Haken-Strobl method [51] were applied to the Hamiltonian matrix containing diagonal site energies and inter-chromophore couplings, for initial excitations occurring on each of the Trp or Tyr residues in the central dimer of the three-tubulin polymer (see Figures 3A and 5). As displayed in Figure 5, strong coherent beating effects—oscillatory behavior indicating that the wavefunction probability is distributed significantly across multiple chromophores—can be observed for 500-600 fs in panels A, C, D, F, and H. Weaker but still appreciable coherence is observed for 200-300 fs in panels B, E, and G. In the infinite-time limit, the Haken-Strobl model converges to equal probabilities for all sites. Previous studies with Trp only [29] showed only minimal coherent effects for the initial pure states described in panels A and E, and oscillatory wavefunction behavior was shared significantly over only two or three sites at a time for all initial pure states. Panels F and H demonstrate strong coherent beating among five chromophores at once, on similar timescales as those observed in the gold-standard Fenna-Matthews-Olson (FMO) and light-harvesting center (LHC) photosynthetic complexes.

The excitation travels from the initial site across the length of the tubulin dimer and beyond, owing to the uninterrupted spacing of chromophores in the polymer. We show in the simulations below that it is feasible for such an arrangement to effectively transfer energy along the protofilament length via a combination of coherent transport and incoherent relaxation/excitation. Additionally, though not considered yet in our model, the unique cylindrical symmetries found in the tubulin chromophore lattices of microtubules may effectively serve to enhance transfer rates and distances, and potentially enable energy transfer along helical pathways.

As 1 ns roughly corresponds to the tryptophan fluorescence time, each of the simulations aggregated in Figure 6A consists of 500 cycles (1999 fs each) of coherent quantum evolution, followed by environmentally induced relocalization on a new tryptophan chromophore based on the quantum probability amplitudes calculated from the Haken-Strobl procedure. Our simulations in Figure 6A ($N$=100) show the stark contrast between the most conservative of scaling factors (Förster) and the more modern descriptions for the protein matrix. Using the Förster approximation generally limits the propagation distance to one tubulin dimer (8 nm) in either direction. By examining the spread of the Förster curves at three dephasing rates, it becomes clear that the single-photon excitation generally travels further with decreased dephasing. The more modern protein descriptions produce much further propagations, going as far as 12 dimers (96 nm) in these simulations. These distances (~0.1 micron) are certainly of physiological relevance, as they span the reach of dendritic length scales and neuronal fine structure. In the Juzeliūnas approximation with a dielectric of 8.41 and a dephasing of 50 cm$^{-1}$ ("Juzeliūnas8.41-50"), more than 80% of simulated photoexcitations were transported two tubulin dimer lengths (16 nm) or beyond, and nearly 40% were transported five tubulins (40 nm) or beyond. There is a conspicuous shift in the mean of the Juzeliūnas curves to the right due to asymmetries in the tryptophan network, and this becomes more pronounced with decreased dephasing. To address the possible objection that using a dielectric constant of 8.41 in the quantum electrodynamical Juzeliūnas description gives a scaling factor exceeding unity, we have also included simulations where the scaling factor is set equal to unity ("Nofactor"). The mean value of tubulin propagation lengths for Juzeliūnas8.41-50 is 3.73, while the mean value for Nofactor8.41-50 is 3.22. For the latter, 79% of photoexcitations were transported two tubulins or beyond.

In a different set of simulations (see Figure 7), the coherent evolution time was shortened by an order of magnitude from 1999 fs to 500 fs (x 2,000 cycles of evolution = 1 ns), for comparison's sake, since the bounds of this evolution time have yet to be determined from experiment. In the Juzeliūnas and Nofactor approximations, the effect on excitonic propagation was clear and pronounced: Mean propagation distances increased across the board, at all dephasing rates. For the Juzeliūnas8.41-50 case, the mean propagation distance increased by 52%, 90% of simulated photoexcitations travelled two tubulins or more, and 65% travelled five tubulins or more. This may be attributed to the strong coherent beating observed on the timescale of each coherent cycle (~500 fs) in five out of eight panels of Figure 5. Sampling these probabilities at earlier stages in the Haken-Strobl evolution (well before the long-time limit), where the excitonic state wavefunction is distributed more on chromophores other than the initially excited pure state, would increase the likelihood of the exciton's localization at a new site. In other words, though stochastic in the localization process of each quantum evolution, the chromophore network conspires in its geometry, coherence times, and quantum transition dipolar properties to enable further exciton movement along the protofilament. This subtle influence by chromophore geometry can have significant consequences at the macroscale, as evidenced most starkly by the conspicuous asymmetric shift toward the right in the tryptophan-only simulations of Figures 6 and 7.

We have further conducted simulations with the addition of the aromatic amino acid tyrosine in our minimal microtubule lattice model. Like tryptophan, tyrosine is also a source of delocalizable electrons but exhibits a transition dipole strength of only about 1.2 debye compared to tryptophan's 6.0 debye. There is, however, a much larger abundance of Tyr (34 sites) than Trp (8 sites) in tubulin, which increases the total number of chromophores to 42.

These simulations suggest that the introduction in our model of Tyr sites, which reduces the average inter-chromophore distance, decreases the mean propagation distance over a range of evolution cycles and dephasing rates (see Figure 6B). This decrease in the mean propagation can be attributed to the greater probability of the excitonic state remaining within a single tubulin, due to the denser packing of proximal chromophores. Still, for Juzeliūnas8.41-50 and Nofactor8.41-50, more than half and nearly a third, respectively, of photoexcitations reached two tubulin lengths or beyond. The impact of the dephasing rate was less pronounced than in the Trp-only simulations, as these fractions stayed constant, decreased slightly, or even increased with greater dephasing. It should be noted here that a wide variety of microtubule decorations, including guanosine phosphates (GDPs, GTPs) and microtubule-associated proteins (MAPs), may also contribute to changing the electromagnetic couplings and energy landscapes for excitonic transport, both along these polymers and, by extension, the cytoskeletal network writ large. Additionally, the electromagnetic fields caused by action potentials induced during diverse global brain states may "tilt" the exciton propagation in one direction, or extend it further in both directions, depending on the specific environmental conditions.

**DISCUSSION**

The most prominent conclusion from our results above is that excitonic propagation via quantum coherent energy transfer in tubulin aromatic networks is at least significant on the micron scale. This conclusion is predicated on (1) the close agreement between tubulin experimental spectra and theoretically predicted curves using the tubulin high-frequency dielectric ($\varepsilon_r$ = 8.41; see Figure 4) and (2) a quantum electrodynamic description of the interstitial protein matrix that modifies the scaling factor for the transition dipole couplings from the traditional Förster type. What this long-range energy propagation suggests is the physiological relevance of coherent dynamics due to photoexcited aromatic amino acid networks, and of the interplay between those coherent dynamics and incoherent "hopping" (i.e., relocalization) events that move the excitonic energy down the length of the tubulin polymer. It also alludes to the possible neuroexcitatory role of UPEs from reactive oxygen species (ROS) produced during normal metabolic processes as well as during periods of intense oxidative stress. These neuroexcitations would be distinct from—but may be involved in signaling or triggering for—the conventional neural firings that are dictated by rates of ion flow through sodium, potassium, and other channels. We have not considered in our simulations the full complement of 13 protofilaments that forms the mammalian microtubule, but in principle the introduction of this more sophisticated aromatic lattice would allow for more topologically intricate excitonic trajectories and possibly new quantum coherent phenomena due to the manifestation of eigenstates with cylindrical symmetry.

A secondary conclusion is the prediction of strong coherent beating effects up to 600 fs (see Figure 5, particularly panels F and H) demonstrating energy transfer among the Trp and Tyr chromophores in tubulin. Such coherent beating is consistent with the lifetimes of oscillatory excitonic behavior in several photosynthetic complexes and would certainly suggest that the experimental search for macroscopic quantum effects in tubulin via modern coherent spectroscopy is reasonably motivated. Experimental confirmation and refinement of the Haken-Strobl site probability evolutions in Figure 5 would allow more precise simulations of the extent of ROS-induced, UPE-stimulated excitonic transport in neuronal microtubules. A key question that remains to be answered is the rate of energy loss with each excitonic relocalization ("hopping") – how much energy is dissipated to the environment and how much remains for functional use in "recoherence" over the aromatic network? Significant dissipation to the environment would severely limit the extent of excitonic energy transported at these length scales; however, some Markovian [52] and non-Markovian [53] models of the environment have actually shown enhanced resonance energy transport due to the system-bath interaction. Future refinements should certainly also address the issues faced with systems of multiple chromophoric types [44], as the introduction of Tyr chromophores posed alignment challenges for the tubulin circular dichroism spectra. We expect that using an extended transition charge density instead of the point-dipole approximation for the inter-chromophore couplings may bring experiment and theory in closer agreement here.

Furthermore, it must be noted that while Onsager/Förster coupling holds in the intermediate range of ~6–20 ångströms, where orbital overlaps can be ignored and only the dipole–dipole interaction is relevant [54, 55], at very small distances (less than ~6 ångströms) orbital overlap-driven mechanisms become important [56] and at very large distances (greater than ~20 ångströms) radiative mechanisms begin to dominate [57]. This last case suggests the use of the Juzeliūnas screening factor in our investigation, since the majority of couplings in our system occur at distances greater than 20 ångströms. As such, the Förster and Juzeliūnas factors are taken as the two extreme cases for screening. A more sophisticated and extended simulation would consider the distance dependence of such screening factors, but this is beyond the scope of the present work.

The mechanism and analysis we propose in this paper for ROS-induced excitonic transport in neurons is generally applicable to all tubulin-based cytoskeletal elements and would be readily adaptable to aromatic networks in other biopolymers. Furthermore, as increased generation of ROS is a hallmark of many diseases as diverse as neurodegeneration, diabetes, and cancer, UPE stimulation of excitonic transport in microtubules may represent a coherent signaling mechanism at the root of multiple aging processes. Supporting this line of thought, recent evidence [58] confirms the existence of "membrane nanotubes" containing F-actin and microtubules, which may serve as a means of long-range cell-to-cell signaling—a kind of intercellular highway. These nanotubes facilitate cell-to-cell communication through a diverse array of methods, ranging from the exchange of a variety of signaling carriers, organelles, and even unicellular organisms, to the enabling of long-distance electrical coupling between cells. As a result of their important role in intercellular signal transduction, membrane nanotubes have been implicated in embryogenesis, differentiation, cellular reprogramming, cancer initiation and progression, and electrophysiological function in neurobiological processes. Intriguingly, the cytoplasm inside membrane nanotubes is densely occupied by mitochondria in a host of instances [58]. It may very well be that tubulin protofilaments in microtubules, along with other biopolymers and perhaps the mitochondrial membrane itself, could be the medium (via aromatic networks) by which optical signals are transmitted coherently between cells. Other researchers have similarly proposed that such long-range signaling should be mediated by optical or photonic channels rather than by the conventional biochemical cascades of the cell, and that myelinated axons may serve as waveguides to couple biophotons and nuclear spins for quantum communications in the brain [59].

The role of the aqueous environment should also be taken into consideration when considering clinical targets for microtubule and neuronal stability in neurodegenerative disease. Several studies by diverse research groups have been undertaken to examine the unique properties of so-called interfacial or "ordered" water, which exists in the hydration shells of biomolecules and exhibits strikingly distinct properties from customary bulk water [60]. In particular, this layer of biological water may be hundreds of microns thick, with a tenfold higher viscosity than bulk water and excluding all solutes. Furthermore, biological water manifests an absorption peak at 270 nm in the UV [61], overlapping with the absorption maxima (265-285 nm) of tryptophan and other aromatic amino acids (i.e., tyrosine and phenylalanine). It then suffices to reason that biological water, akin to a waveguide, may serve as an "insulator" for coherent propagation of energy via aromatic networks in microtubules, which are literally surrounded by ordered biological water. Thus, instead of emitting as UPEs, these photons would remain within the microtubule system for cytoskeletal signaling and energy transport. Indeed, the potential for such laser-like behavior emerging from the interaction between water electric dipoles and the biological radiation field has been studied in the formalism of quantum electrodynamics since at least the 1980s [62]. More recent efforts in this vein have focused on biologically specific instantiations of the formalism to study the interactions between DNA and enzyme systems [63]. The generality of this description of collective dipole modes [64] for constituent aromatic rings of biomacromolecules ensures that such a framework could also be applied to microtubule aromatic networks, and indeed, theoretical studies of anesthetic alterations to these collective dipole modes in tubulin are beginning to bear fruit [65]. While outside the scope of our present study, a truer picture of excitonic transport due to ROS-generated UPEs should include the effects from a realistic interfacial water field.

**Potential Implications for Tauopathic Diseases:**

Specific protein inclusions define most neurodegenerative pathologies – for example, neuritic plaques and neurofibrillary tangles (NFTs) in Alzheimer's disease (AD), argyrophilic inclusions (Pick bodies) in Pick's disease, and Lewy bodies and Lewy neurites in Parkinson's disease. It was established in the 1980s and early 1990s that paired helical and straight filaments in the brains of AD patients are made of isoforms of microtubule-associated protein (MAP) tau, in a hyperphosphorylated state. Fronto-temporal lobar degeneration with tau inclusions, progressive supranuclear palsy, corticobasal degeneration, and Parkinson's disease-related dementia have all shown similar pathology. Most recently, chronic traumatic encephalopathy, a neurodegenerative disease resulting from environmental assaults including repetitive blast or concussive injuries, has been characterized by filamentous tau inclusions [9].

As the conversion of soluble to insoluble filamentous tau protein is central to many human neurodegenerative diseases, tau pathways are excellent potential targets for therapeutic effect. Tau represents a family of isoforms that differ from each other by the presence or absence of amino acid repeats in the amino-terminal and carboxyl-terminal domains of the protein [66]. These repeats and some adjoining sequences constitute the microtubule-binding domains of tau. In AD, after tangle-bearing cells die, tau filaments can remain in the extracellular space as so-called "ghost tangles" consisting largely of the repeat region of tau, and experiments have shown that aggregates made of the repeat region can induce neurotoxicity. However, in other human tauopathies, such as Pick's disease, progressive supranuclear palsy, corticobasal degeneration, and most cases caused by tau mutations, filamentous tau does not accumulate in the extracellular space after the death of aggregate-bearing nerve cells. The reasons underlying these differences remain to be established.

In many diseases in which tau filaments are implicated, tau is unable to interact with microtubules. Currently, microtubule-stabilizing agents such as taxanes and epithilones are actively being pursued as treatments for such diseases [1-4]; however, crossing the blood-brain barrier remains an issue for large compounds. Moreover, there is convincing evidence for the role of oxidative stress in the pathogenesis of chronic diseases, particularly in neurodegenerative disorders where misassembly and aggregation of proteins results due to impaired or insufficient cellular defenses. The manifestation of misfolded proteins and aggregates is a hallmark of AD, Parkinson's disease, amylotrophic lateral sclerosis, polyglutamine diseases, diabetes, cancer, and many others, each of which exhibits age-dependent onset.

**The role of amyloid-β:** The amyloid cascade hypothesis of AD [67] postulates that the accumulation of amyloid-β (Aβ) peptide fragments leads to Aβ oligomerization, tau hyperphosphorylation and aggregation, synaptic dysfunction, nerve cell death, brain shrinkage, and ultimately cognitive decline. Oxidative stress (excess ROS), excitotoxic mechanisms, and overactivity of some brain networks seem to be important in the coordination of Aβ's toxic effects. However, there is considerable debate about the role that Aβ plays in AD pathogenesis [68] because, in fact, a substantial fraction of cognitively healthy older people have Aβ plaques that would satisfy commonly applied AD diagnosis criteria [69]. Interestingly, a modified version of the amyloid cascade hypothesis maintains that Aβ by itself is not dangerous, but rather that Aβ produces oxidative stress and AD in combination and in parallel with the dyshomeostasis of metal ions such as copper, iron, and zinc [70-71].

The relationship between Aβ and tau remains to be fully elucidated. It is clear, though, that their roles in the sequential events of AD progression are deeply interrelated. While neuron hyperexcitability, which occurs after Aβ exposure, is reduced in the absence of tau, studies investigating the appearance of Aβ deposits and tau aggregates in the human brain as a function of age suggest that tau inclusions appear at a younger age than do Aβ plaques.

Indeed, tau hyperphosphorylation seems to be an early and crucial event in tau-mediated neurodegeneration. A healthy human brain has on average 1.9 moles of phosphate per mole of tau, whereas tau from abnormal filaments of patients with AD carries 3-4 times as much phosphate per mole of tau. Many tau phosphorylation sites are known; so too are candidate protein kinases and phosphatases. The amount of phosphorylation depends on the conformation of tau and on the balance between the activities of tau kinases and tau phosphatases. Inhibition of tau kinases and activation of tau phosphatases are therapeutic targets for AD and other tauopathies. However, because these enzymes have several substrates, whether effective and safe tau kinase inhibitors and tau phosphatase activators can be developed remains to be seen. Hyperphosphorylation of tau reduces its ability to interact with microtubules, suggesting a partial loss of stability and function that might be triggering the toxic function mechanisms of disease. This observation implies that microtubule stabilization could be of therapeutic benefit, which is the insight that has motivated our present study into photoexcitatory mechanisms in microtubule networks.

Terahertz beating of collective dipole modes for the delocalized electrons of tubulin aromatic amino acids [65] may play a critical role in the fractal hierarchy of oscillations in the cell and organism at gigahertz (tubulin network), megahertz (microtubule resonance), kilohertz (neural firing), and hertz (brainwave gamma synchrony) frequencies. To speculate on the purpose of such a hierarchy, terahertz oscillations in tubulin may act as a fine-scale clocking mechanism affecting overall neuron function [72], neuronal synchrony [73], and electroencephalogram readings [74]. Impairment of this mechanism underlying synchrony could also explain loss of cognitive abilities. Degradation of microtubule networks would disrupt global synchronies and/or coherences between neurons and distant portions of the brain, which may be critical in the generation and reproduction of memory [75-76]. Thus, stability and vibrational spectra of microtubules may prove to be essential markers of mental health and cognitive function (e.g., in AD, post-operative cognitive dysfunction, traumatic brain injury, depression, and stress disorders).

How does our modeling of microtubule aromatic networks relate to tauopathies? While the lack of a crystal structure for MAP tau limits the applicability of our methods to its aromatic amino acids, the canonical tau isoform has 0% Trp and ~0.8% Tyr in the total sequence, considerably lower than the percentages in tubulin. MAP tau may thus serve as a potential "sink" for UPEs, with photoexcitation inducing subtle structural changes in the cytoskeletal matrix that hasten or retard microtubule degradation, thus altering the rate of accumulation of tau pathologies. Unlike tau, MAP2 (found primarily in dendrites and implicated in learning and memory) possesses 0.3% Trp and 1.4% Tyr, with the first 305 amino acids at the N-terminal side of the protein containing 1.3% Trp, potentially facilitating exciton transport through microtubules or acting as a bridge between adjacent microtubules, helping to further transfer or dissipate ROS-generated UPE energy. NFT formation leading to tau overexpression and decreased MAP2 in dendrites [77-81] can therefore lead to altered energy absorption profiles that may debilitate neuronal function, which is consistent with the findings of other approaches.

**Potential for Using Ultraweak Photon Emissions in Diagnosis:**

The nature of ultraweak photon emission (UPE) in biological systems has been reviewed by several authors previously [13, 82-86]. The advent of new photon-counting technologies in the early 1960s provided the tools to demonstrate the existence of low-level luminescence in all living organisms, to be distinguished from the strong bioluminescence of, for example, various firefly and squid species. Several reports have suggested that the effects of electromagnetic signaling in the cell have the characteristics of receptor-mediated interactions.

In a series of studies spanning a period of some 25 years [87-94], G. Albrecht-Buehler demonstrated that living cells possess a spatial orientation mechanism located in the centrosome [91-93]. This is based on an intricate arrangement of microtubules in two perpendicular sets of nine triplets (called centrioles). This arrangement provides the cell with a primitive "eye" that allows it to locate the position of other cells within a two-to-three-degree accuracy in the azimuthal plane and with respect to the axis perpendicular to it [93]. He further showed that electromagnetic signals are the triggers for the cells' repositioning and observed that cultured cells move toward infrared (IR) light [91]. It is still largely a mystery how the reception of electromagnetic radiation is accomplished by the centrosome, but Albrecht-Buehler proposed that centrosomes are IR detectors and that microtubules are cables carrying signals between subcellular organelles.

The finding of unique patterns of UPE emitted from biological systems during specific phases of the cell cycle [15] suggests that UPE is specific and that UPE could have a role in non-chemical, distant cellular interactions [95], rather than being a random or isolated event.

Volodyaev and Beloussov [18] regard the UV component of UPE as "proven" but distinctly different in source from visible UPE, based on the experimental evidences of Tilbury and Quickenden [96], Gurwitsch et al. [97], and Troitskii et al. [98], and note the coincidence of UV UPE with the historical mitogenetic effect—a change in mitotic regime in a cell culture or tissue under the non-chemical influence of an external biological object—in both spectral range and the culture growth phase, when they are observed. In an extensive work, the group of Konev [15] detected UPE from several dozens of various species, ranging from bacteria to vertebrates and higher plants. The mechanism of the UV emission was not discovered in detail, but it was shown different from lipid peroxidation, and supposedly connected to protein synthesis [99]. The known inductors of the mitogenetic effect included actively growing microbial and tissue cultures; working muscles and heart; excited neurons; blood of healthy people; malignant tumors; resorbed and regenerating tissues; and certain chemical reactions [18]. A substantial bulk of the mitogenetic investigations dealt with neural excitation and brain tissue activity, with authors showing propagation of mitogenetic activity along the excited nerve fiber and reporting that mitogenetic spectra of nerves depended on the nature of exciting agents. In another series of experiments, flashes of photon emission (called "degradational radiation") were detected immediately after application of stressful agents [100]. Kaznacheev et al. [101-102] observed optical biocommunication between cell cultures that were completely shielded from the environment except for optical coupling via a thin quartz window—transparent to UV wavelengths—between them. UV photons emitted from irradiated cells appear to have been absorbed by cells in the unirradiated culture and are thought to have caused their death within 12 hours. These results have been achieved and replicated using unicellular green alga, immortalized mouse fibroblasts, and adult human microvascular endothelial cells.

The question of UPE signaling (i.e., mechanisms of distant communication of biological systems) was revisited by Fritz-Albert Popp in the 1980s [103]. His main hypothesis was that biological systems possessed an inner coherent electromagnetic field. Interestingly, Popp et al. [104] discovered a wide spectrum of UPEs from living cells in the 200-800 nm range, and his group suggested that such biophoton emissions were coherent and seem to originate from DNA of the cell nucleus. Thus, they could be detected by other coherent-state systems. More recently, Kurian et al. have calculated that London-force dipole coupling between DNA base pairs can produce coherent oscillations of physiological significance [64]. These coherences may be protected from thermal buffeting by the formation of tightly bound DNA-protein complexes that exclude water and counterions from the DNA surface. With possible detection decades earlier by Popp and colleagues, Ref. [64] has proposed that such quantized oscillations in the terahertz regime may be a source of coherent energy for DNA metabolism, chromosomal organization in mitosis, and meiotic recombination.

The emission of photons from cells is typically very weak, on the order of 10 - 10,000 photons/s/cm$^2$ from the cell surface, as demonstrated in bacteria, yeast, whole animals, and plants, as well as cells and homogenates from organisms [105]. However, these UPEs are not negligible as they are currently being exploited as a diagnostic tool to monitor oxidative stress [105]. Whereas most of the techniques to detect and measure ROS are invasive and require the destruction of living structures, a number of researchers have drawn attention to the measurement of this low-level chemiluminescence for detecting electronically excited states in biology [105-106]. Such photon production has been related to the direct utilization of molecular oxygen and the oxygen-dependent chain reactions involved in lipid peroxidation processes [107-108]. Since the early 1980s, it has been known that peroxidal lipid reactions are important components in the etiology of diabetes, liver and lung diseases, arteriosclerosis, aging, and cancer [109-113]. The production of oxygen radicals during the respiratory burst of phagocytic cell activity plays an essential role in bacterial killing and in regulating the processes of acute inflammation [114-115]. This realization led to the idea that UPE measured in blood might be a general marker for health and could be used for diagnostic and clinical purposes [116-117]. Early studies established the interdependence between various diseases and UPE intensity by measuring the differences between the luminescence of the blood of healthy and diseased human subjects. UPE of blood from patients with diabetes mellitus, carcinomas, and hyperlipidemia showed higher emission levels than those of the samples from healthy controls [116]. Human UPE has also been characterized anatomically [117-118]. Whether UPE has a functional biological purpose, or may just represent excess "waste" energy, is the current area of investigation.

Based on the available evidence, the belief that cellular signaling with light is either impossible or not achievable under physiological conditions is exceedingly tenuous. Such disbelief is consistent with mainstream biophysics, which considers UPE from cells to be random photon emissions due to cellular metabolism, and is also consistent with the weak emissions measured from the cell surface. However, the real biophoton intensity within cells can be significantly higher [103, 119] than the one expected from UPEs usually measured some centimeters away from the cells, based on the $1/r^2$ dependence of the intensity. Mitochondria are the cell's primary producers of ROS [120-121] (see also Figure 2), and therefore are expected to mark the epicenter of UPE emission in the cell. Estimates indicate that internally at least $10^8 - 10^9$ photons per second can be produced in mitochondria-rich retinotopic visual neurons [119]. This back-of-the-envelope calculation may also prove to be an underestimate, for two reasons: (1) photons can be absorbed and stored in coherent domains of ordered water [62], never to be emitted in the timescale of a single experiment, and (2) photons can be coherently transported and dissipated along biopolymers (e.g., microtubules) via aromatic networks, to be used for cellular signaling. These ubiquitous UPE absorbers in the cell may also explain the discrepancy of four to eight orders of magnitude between the high production of intracellular biophotons and the low count numbers detected from the cell surface. Alternatively, the electron transport chains on the inner membrane of mitochondria contain photosensitive chromophores such as flavinic, pyridinic, and porphyrin rings [20, 119], which could serve as conduits for UPE energy. Fluorescent lipid chromophores—formed during regulated lipid peroxidation of membranes—can also act as photoacceptors [119].

This process is of particular importance to neuronal microtubules as they co-localize with filamentous mitochondria, which are typically 100–500 nm in diameter and up to 10 microns in length, forming complex reticula (networks) [20]. Rough calculations [20] further suggest that filamentous mitochondria and microtubules within neurons can act as optical waveguides for biophotons at wavelengths greater than a cut-off of 145 nm, which would comprise the UV, visible, and IR. Based on the results obtained by different authors [122], it can be concluded that the major part of UPE is in the visible and IR regions of the spectrum, but the contribution in the UV region cannot be completely ruled out [83]. Indeed, our analysis has shown that excitonic propagation from singlet states in aromatic networks is stimulated by UV photoemissions and may be relevant for coherent signaling in the brain, at least on dendritic length scales. Analogously, triplet excitations of aromatic amino acids by UPE in the visible and IR regions may provide an additional mode of coherent transport for these longer wavelength emissions.

Feasible mechanisms for UV-induced changes in microtubule architecture include the reduction of disulfide or peptide bonds stimulated by photoexcitation of tryptophan or tyrosine groups, or subtle protein structural changes owing to alterations in aromatic flexibility. Functional microtubule networks may use the energy from this UV light as a signaling mechanism throughout the cytoskeleton. Microtubule networks in cognitively impaired and Alzheimer's patients may be unable to support effective channeling of these UV photons for signaling, resulting in a UV emission surplus, and thus leading to cellular oxidative damage and hastened aging.

**METHODS**

We have developed computational models of aromatic amino acid networks in tubulin, which comprise the building blocks of neuronal microtubules that are implicated in many tauopathic disorders, including Alzheimer's and Parkinson's diseases. To robustly capture macroscopic, neuronal-level effects from electronic oscillations at the molecular level, we modeled tubulin aromatic networks from first-principles density functional theory methods and fully quantum transition dipole coupings, and then connected those results to protein- and polymer-level dynamics derived from classical or semi-classical approaches. The physical intuition for this process is visualized in Figures 2 and 3. To analyze the impact of UV photoexcitation on tubulin networks, we applied techniques from Förster resonance energy transfer, which remains the dominant theory applied in electronic energy transport to describe the coherent conveyance of electronic energy across chromophores via electrodynamic coupling of their transition electric dipole moments [40]. To compute excitonic coupling interactions for comparison with experimental data, we implemented the Schrödinger equation with a tight-binding Hamiltonian for an interacting $N$-body system. Here, we give a brief summary of the computational procedures yielding site energies, excitonic couplings, optical spectra, and time evolution of exciton populations. All data will be made available upon reasonable request.

**Molecular Dynamics (MD) Simulation/ Dominant Tryptophan Conformation:** MD simulations were performed in Ref. [29] based on previous methods [123-124], using NAMD [125]. Coordinates for the $\alpha\beta$-tubulin dimer at 3.5 Å resolution, GDP, GTP, and magnesium ion are from the Brookhaven National Laboratory Protein Data Bank [126] entry 1JFF [127], and PDB2PQR 1.7 [128] was used to generate per-atom charge and radius at pH 7.0 using PROPKA [129-130] and the AMBER [131] force field. Swiss-PdbViewer 4.0 [132] was used to model the missing residues in the crystal structure ($\alpha$-tubulin: 1, 35-60, 440-451, $\beta$-tubulin: 1, 438-455). GDP and GTP were modeled for use by the AMBER 94/99 force field by Meagher et al. [133]. PTRAJ, using the AMBER99SB force field [134] from AMBER 10 [131] was used to neutralize the system, add counterions to reproduce physiological ionic concentrations, and add a TIP3P water cube buffer of 25 Å (63,283 molecules of water). After minimization and heating to 310 K, equilibration occurred for 519 picoseconds (ps) while gradually releasing constraints on backbone atoms. Then, using periodic boundary conditions, a 17-nanosecond (ns) simulation occurred where atomic coordinates were saved from the trajectory every 2 ps.

PTRAJ was used to analyze the 17-ns trajectory for root mean square deviation (RMSD) of the backbone of non-C-terminus residues of both α- and β-tubulin, as well as the β-factor of all residues. The RMSD was found to stabilize after approximately 8 ns. The final 8 ns of the total trajectory were chosen for further analysis. The GROMACS program g_cluster [135] was used to cluster the Trp atom positions of all frames via single linkage with a cut-off of 0.5 Å. Trp was chosen because it is the dominant absorber of UV wavelengths and is thus expected to affect the overall absorption spectrum the most. The middle structure of the largest cluster was taken as the dominant conformation for the Trp residues. All other atom positions (e.g., in Tyr residues) were taken from the frame of the MD trajectory in which the dominant confirmation of Trp residues was found. In this way, the Trp residues serve as anchor points for the rest of the protein.

**Excitation Energies for Chromophore Complex / Site Energy Calculations:** Site energies for tubulin chromophores were derived from structure-based calculations of the free energy change that occurs when the ground-state charge density of Trp or Tyr $m$ is shifted to the first excited state, analogous to methods used for the protein-pigment complex (PPC) in FMO [136-137]. Quantum chemical calculations of the chromophores *in vacuo* yield the charge distributions of the $S_0$ and $^1L_a$ / $^1L_b$ states and a contribution $\Delta E_{qm}$ to the $S_0 \rightarrow {}^1L_a$ or $S_0 \rightarrow {}^1L_b$ transition energy due to the differing orientations of the chromophores. Trp and Tyr side chains were each isolated from the dominant conformation (described above) and capped with a hydrogen atom to complete the valence. *Ab initio* geometry optimizations at the ground state were performed using density functional theory (DFT) at the B3LYP/6-311G(d,p) [138] level. Quantum chemical calculations of the eight lowest singlet excitation energies of the Trp network in tubulin were done using time-dependent density functional theory (TDDFT) [139-140] at the B3LYP/6-311+G(d,p) [138] level. All calculations were performed using ORCA v2.8 (http://cec.mpg.de/forum/).

The transition energies for the $^1L_a$ state of Trp and the $^1L_b$ state of Tyr were found to be comparable to previous work using similar methods [141]. However, like these results, our values (37660 cm$^{-1}$, average of the Trp transition energies) were found to be approximately 2% lower than experimentally determined values (38473 cm$^{-1}$) [141-142]. Because this 2% difference corresponds to less than one cm$^{-1}$ difference in $\Delta E_{qm}$, this discrepancy is negligible in our model results. The relative quantum correction $\Delta E_{qm}$ to the site energy was thus taken as the difference between the calculated TDDFT and experimental values.

The second part of the site energy calculation results in a contribution $\Delta E_{coul}$ due to classical electrostatic interaction between the charge distributions of the $S_0$ and $^1L_a$ states, or $S_0$ and $^1L_b$ states, and the protein environment. Using the charge density coupling method [136-137], the site energy shift of the $m^{th}$ chromophore is calculated from the Coulomb interaction of the difference of the $S_0$ and $^1L_a$ / $^1L_b$ state partial charges $\Delta q_i^{(m)}$, obtained from the Löwdin atomic point charges in the above quantum chemical calculations, with the remaining background charges of the protein $q_j^{(bg)}$. Though there can be large uncertainties when representing the chromophore charge distribution as Löwdin atomic point charges, we have chosen to leave these charges empirically unscaled, instead of artificially improving alignment with experimental spectra through a scaling factor. The electrochromic shift $\Delta E_{coul}$ is

$$\Delta E_{coul} = \frac{1}{\varepsilon} \sum_{i=1}^{N} \sum_{j=1}^{K} \frac{\Delta q_i^{(m)} \cdot q_j^{(bg)}}{\left| r_i^{(m)} - r_j^{(bg)} \right|} ,$$

where $N$ is the total number of partial charges of the $m^{th}$ chromophore, $K$ is the total number of background partial charges (including other chromophore residues), and $|r_i^{(m)}-r_j^{(bg)}|$ is the distance between the $i^{th}$ partial charge of the $m^{th}$ chromophore and the $j^{th}$ partial charge of the background. Background charges were taken from the AMBER99SB force field [134]. Because the measured high-frequency dielectric for tubulin is currently under debate [37], we choose a dielectric value of 2.0, 4.0, or 8.41 to span the entire possible range of expected and measured values [36]. The site energy for the $m^{th}$ chromophore is thus given by

$$E_m = E_0 + \Delta E_{qm} + \Delta E_{coul},$$

where $E_0$ is taken as 35088 cm$^{-1}$ (285 nm) for Trp and 35714 cm$^{-1}$ (280 nm) for Tyr to align with experimental spectra (described below).

**Transition Dipole Coupling Interactions:** Excitonic interactions are calculated using the point-dipole approximation, valid when the distance between chromophores is large compared to the extent of their transition charge densities. The transition dipole couplings are thus given by

$$V_{mn} = f \frac{\mu_m^{vac} \mu_n^{vac}}{R_{mn}^3} \left[ \hat{e}_m \cdot \hat{e}_n - 3(\hat{e}_m \cdot \hat{e}_{mn})(\hat{e}_n \cdot \hat{e}_{mn}) \right],$$

where $\hat{e}_m$ is a unit vector along the transition dipole moment of the $m^{th}$ chromophore, the unit vector $\hat{e}_{mn}$ is oriented along the line connecting the centers of chromophores $m$ and $n$, $\mu_m^{vac}$ is the transition dipole moment of the $m$th chromophore in vacuum, and the factor $f$ effectively takes into consideration screening and local field effects. In the simplest approximation due to Förster, $f = 1/\varepsilon_r$, accounting only for screening effects. However, when screening effects of the medium as well as local field effects are included, this becomes $f = \varepsilon_r^{-1}[(\varepsilon_r+2)/3]^2$ [41-43]. Here we consider both these cases. The optical dielectric constant for a typical protein environment is generally in the range of 2.0 – 4.0, but tubulin's high-frequency dielectric coefficient has been measured as 8.41 [36]. For comparison, we consider the extrema at 2.0, 4.0, and 8.41.

**Alignment with Experimental Spectra:** The tight-binding Hamiltonian for an interacting $N$-body system in the presence of a single excitation is given by [143]

$$H_S = \sum_{m=1}^{N} E_m |m\rangle\langle m| + \sum_{m<n}^{N} V_{mn} (|m\rangle\langle n| + |n\rangle\langle m|),$$

where the states $|m\rangle$ denote the excitation being at site $m$. The site energies and coupling terms are given by $E_m$ and $V_{mn}$, respectively, and are calculated from Coulombic and dipole-dipole interactions, respectively, as described above. Exciton stick spectra for absorption, circular dichroism (CD), and linear dichroism (LD) were calculated from the orientations of the Trp and Tyr molecules in the dominant confirmation and the diagonalized Hamiltonian matrix, using the formulae described by Pearlstein [44]. Theoretical spectra were calculated by applying a Gaussian function to each exciton transition using an assumed full-width half maximum value of 3,500 cm$^{-1}$, which is commonly observed for Trp [144-145]. These were then compared to experimental absorption, CD, and LD spectra adapted from Mozzo-Villarías et al. [146], Clark et al. [147], and Marrington et al. [148], respectively.

**Exciton Population Dynamics:** Time evolutions of the exciton population in the chromophore network in the presence of thermal fluctuations of the environment were modeled using the method pioneered by Haken and Strobl [51]. Here it is assumed that thermal fluctuations of the environment couple to the chromophores by an exciton-phonon interaction, acting only on the diagonal elements of the Hamiltonian system $H_S$, with the fluctuations being unbiased, uncorrelated, and Gaussian in nature. Under these assumptions, the Markovian quantum master equation (known as the Lindblad equation) for the density operator $\rho$ in the Schrödinger picture is

$$\dot{\rho}(t) = -\frac{i}{\hbar}[H_S, \rho(t)] + L_\phi(\rho(t)),$$

where the pure-dephasing Lindblad super-operator (i.e., the *dissipator*) is given by

$$L_\phi(\rho(t)) = \sum_m \gamma_m \left[ A_m \rho(t) A_m^\dagger - \frac{1}{2} A_m^\dagger A_m \rho(t) - \frac{1}{2} \rho(t) A_m^\dagger A_m \right],$$

with $A_m$ and $A_m^\dagger$ being the creation and annihilation (collapse) operators for the exciton-phonon interaction through which the environment couples to the system, and $\gamma_m$ the pure dephasing rate for the $m^{th}$ chromophore. As a reasonable assumption, we take $\gamma_m$ to be 50, 150, or 300 cm$^{-1}$ for all $N$ sites. Population dynamics were computed using QuTip 2 [149].

**Exciton Propagation Simulations**: Exciton evolutions from the Haken-Strobl model for the aromatic network provide the quantum trajectory of site probabilities for energy transfer from the initially excited chromophore. The aromatic network in the linear tubulin polymer is simulated by a minimal system of three identical tubulins, shifted by multiples of 81.9 ångströms [47-49] along the protofilament axis using PyMOL v1.8.2.3 [50], with excitonic jumps to neighboring tubulins always calculated from quantum probabilities in the central tubulin in the triplet, thereby assuming equivalence of tubulins in an infinitely long protofilament. These probabilities are sampled over the entire network at a regular interval (less than 2000 femtoseconds) for each simulation, with repeated cycles of coherent Haken-Strobl evolution followed by relocalization of the exciton on a single chromophore in the network based on the quantum probabilities. These simulated cycles continue for the duration of the fluorescence time of tryptophan (approximately 1 nanosecond), with the intermediate locations of the exciton recorded at each interval timestep throughout the simulation.

**ACKNOWLEDGMENTS**

PK was supported for this project by the National Center For Advancing Translational Sciences of the National Institutes of Health (NIH) under Award Number TL1TR001431. The content of research reported in this publication is solely the responsibility of the authors and does not necessarily represent the official views of the NIH. PK would also like to acknowledge partial financial support from the Whole Genome Science Foundation. TOO is supported by the National Institute on Aging at the NIH (grants R01 5R01AG31517, 5R01AG045058 to Obisesan TO). TJAC would like to acknowledge financial support from the Department of Psychology and Neuroscience and the Institute for Neuro-Immune Medicine at Nova Southeastern University (NSU), and work in conjunction with the NSU President's Faculty Research and Development Grant (PFRDG) program under PFRDG 335426 (Craddock – PI).

**FIGURE CAPTIONS**

Figure 1. Schematic illustration of microtubule depolymerization in Alzheimer's disease neurons. (A) Microtubule (gray) stabilized by tau (red). (B) Destabilized microtubule and free tau with additional phosphorous (yellow). (C) Neurofibrillary tangles formed from phosphorylated tau. This image is a reproduction of the original found in Ref. [70] and presented under the PLoS One Creative Commons Attribution 4.0 International Public License (CC BY 4.0).

Figure 2. Coherent energy transfer in microtubule chromophore networks is stimulated by ultraweak photoemissions due to mitochondrial reactive oxygen species (ROS) production. Filamentous mitochondria are co-located with microtubules in the brain, suggesting that mitochondrial ROS production during respiratory activity may affect neuronal activity. Specific ROS (red and white), particularly triplet carbonyls (red and black), emit in the UV range, where aromatic networks composed of mainly tryptophan and tyrosine may be able to absorb and transfer this energy along the length of neuronal microtubules. The propagation of these excitons extends on the order of dendritic length scales and beyond, indicating that ultraweak photoemissions may be a diagnostic hallmark for neurodegenerative disease and have implications for aging processes.

Figure 3. Arrangement of 42 chromophores – eight tryptophan (W) and 34 tyrosine (Y) – in one tubulin αβ dimer, constructed with PyMOL [50]. Dimer is laterally oriented with the microtubule protofilament axis along the horizontal, microtubule surface above, and microtubule lumen below. C-terminal tails not shown. (A) Location within the tubulin dimer. (B) A representative sampling of the center-to-center spacings between chromophores, in ångströms.

Figure 4. Alignment of calculated spectra with experiment. Experimental spectra (black) and calculated spectra (red, yellow, blue) derived from applied Gaussian broadening of bare stick spectra at 3500 cm$^{-1}$ full-width half maximum for absorption, circular dichroism (CD), and linear dichroism (LD) of tubulin at 300 K. Calculated spectra are shown for (A) *tryptophan-only* and (B) *tryptophan and tyrosine* aromatic lattices. All spectra are plotted in scaled units similar to the Vulto *et al*. treatment [150].

Figure 5. Time evolution of exciton states in tubulin at 50 cm$^{-1}$ dephasing in the Juzeliūnas approximation with $\varepsilon_r$ = 8.41, calculated from the Haken-Strobl model. Tryptophan (Trp) and tyrosine (Tyr) constitute the chromophore network, though only initial excitations on Trps are shown. Initial pure states are set with probability 1 on (A) Trp 1 (αW-21), (B) Trp 2 (αW-346), (C) Trp 3 (αW-388), (D) Trp 4 (αW-407), (E) Trp 5 (βW-21), (F) Trp 6 (βW-103), (G) Trp 7 (βW-346), and (H) Trp 8 (βW-407). Trp locations in parentheses are defined within the tubulin dimer in Figure 3.

Figure 6. Simulations (*N*=100) of coherent energy transport (1 ns each, 500 evolutions x 1999 fs timesteps, with $\varepsilon_r$ = 8.41 and 50-300 cm$^{-1}$ dephasing) due to single-photon excitations of aromatic amino acid chromophore lattices in linear polymers of tubulin dimers. Initial photoexcitation impinges on tubulin 0 at Trp βW-346, which is closest to the protofilament surface. Final propagation distances for all 100 simulations have been tabulated and are shown on the vertical axis. The tubulin dimer on which the exciton finally lands after 1 ns is plotted on the horizontal axis. (A) Exciton propagation in *tryptophan-only* lattices of linear tubulin polymers. (B) Exciton propagation in *tryptophan and tyrosine* lattices of linear tubulin polymers.

Figure 7. Simulations (*N*=100) of coherent energy transport (1 ns each, 2000 evolutions x 500 fs timesteps, with $\varepsilon_r$ = 8.41 and 50-300 cm$^{-1}$ dephasing) due to single-photon excitations of *tryptophan-only* lattices in linear polymers of tubulin dimers. Initial photoexcitation impinges on tubulin 0 at Trp βW-346, which is closest to the protofilament surface. Final propagation distances for all 100 simulations have been tabulated and are shown on the vertical axis. The tubulin dimer on which the exciton finally lands after 1 ns is plotted on the horizontal axis.

**FIGURES**

Figure 1

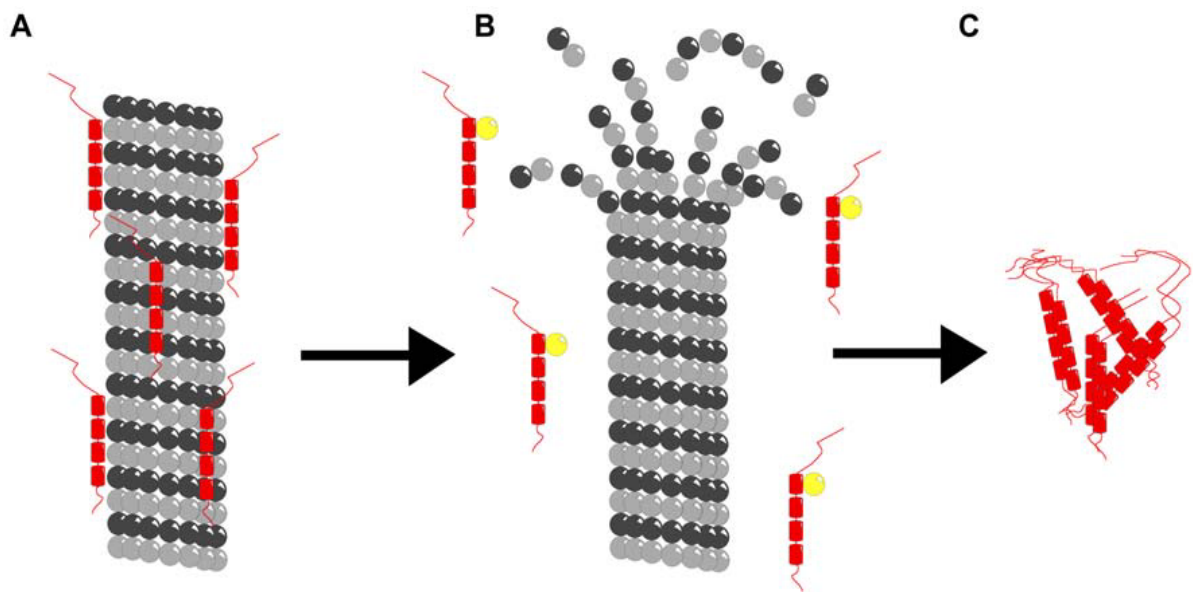

Figure 2

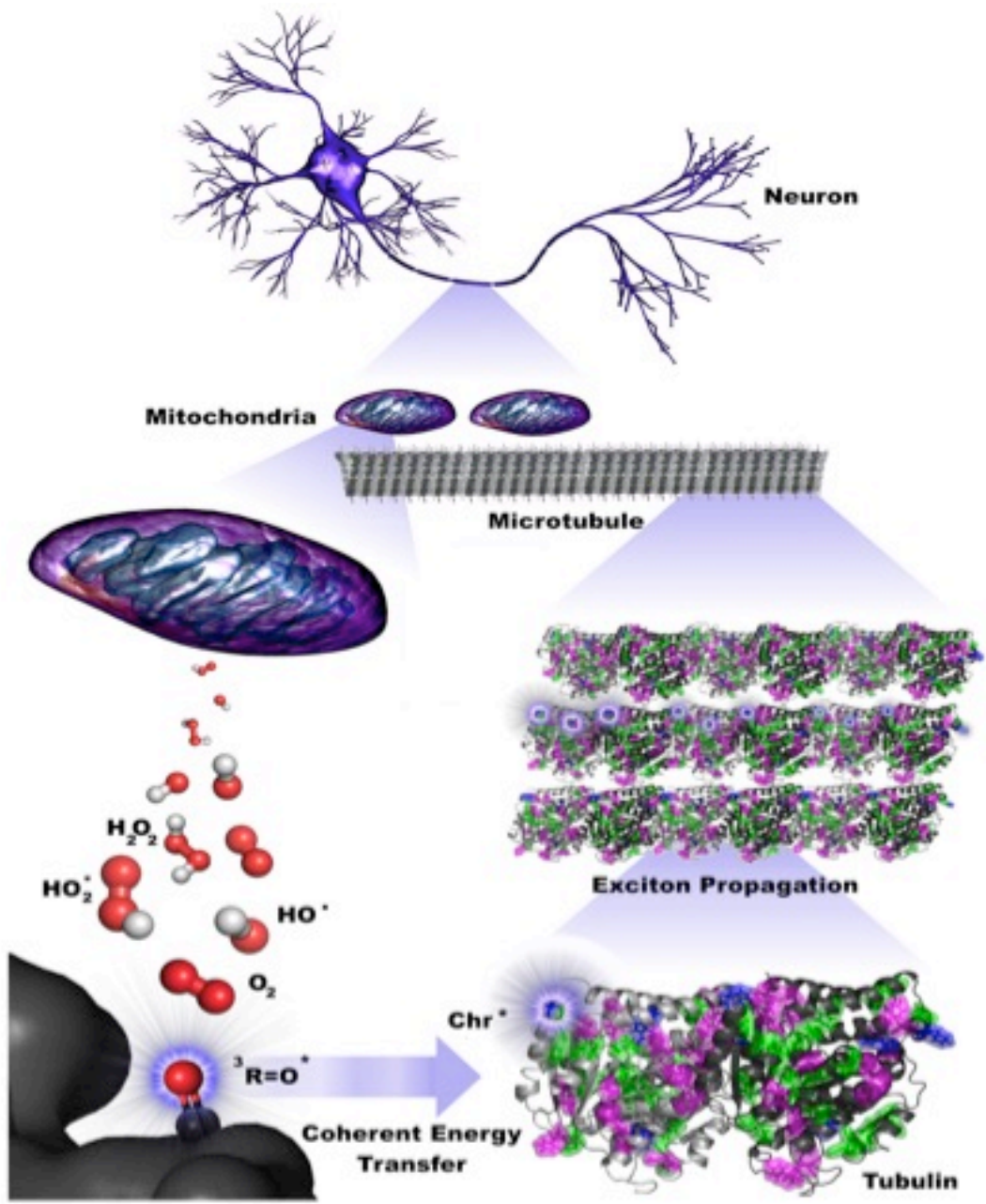

Figure 3

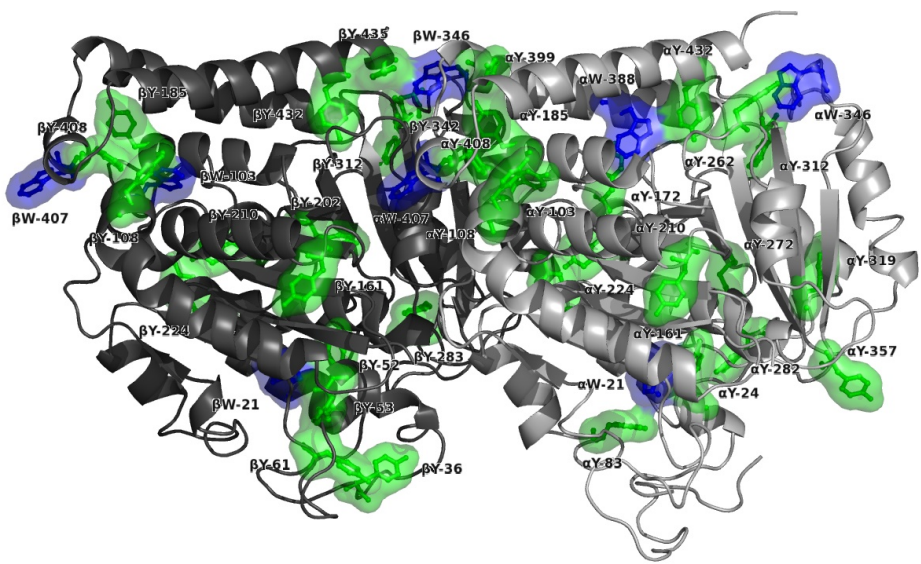

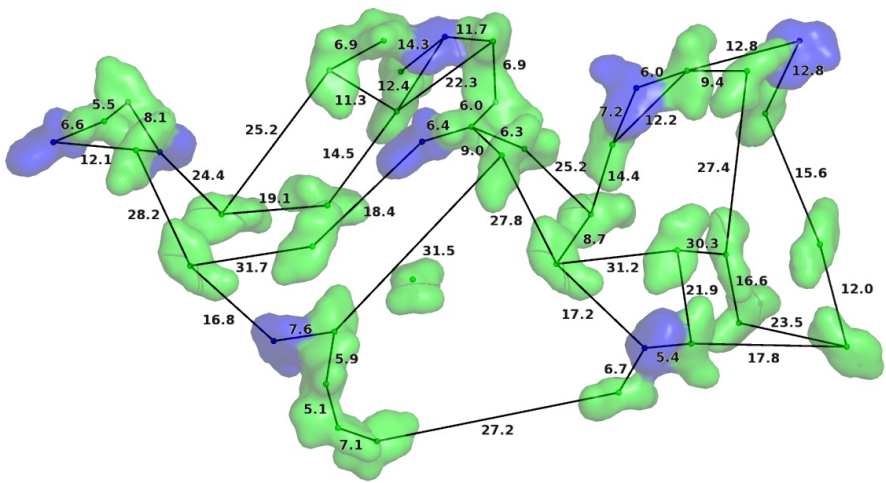

Figure 4

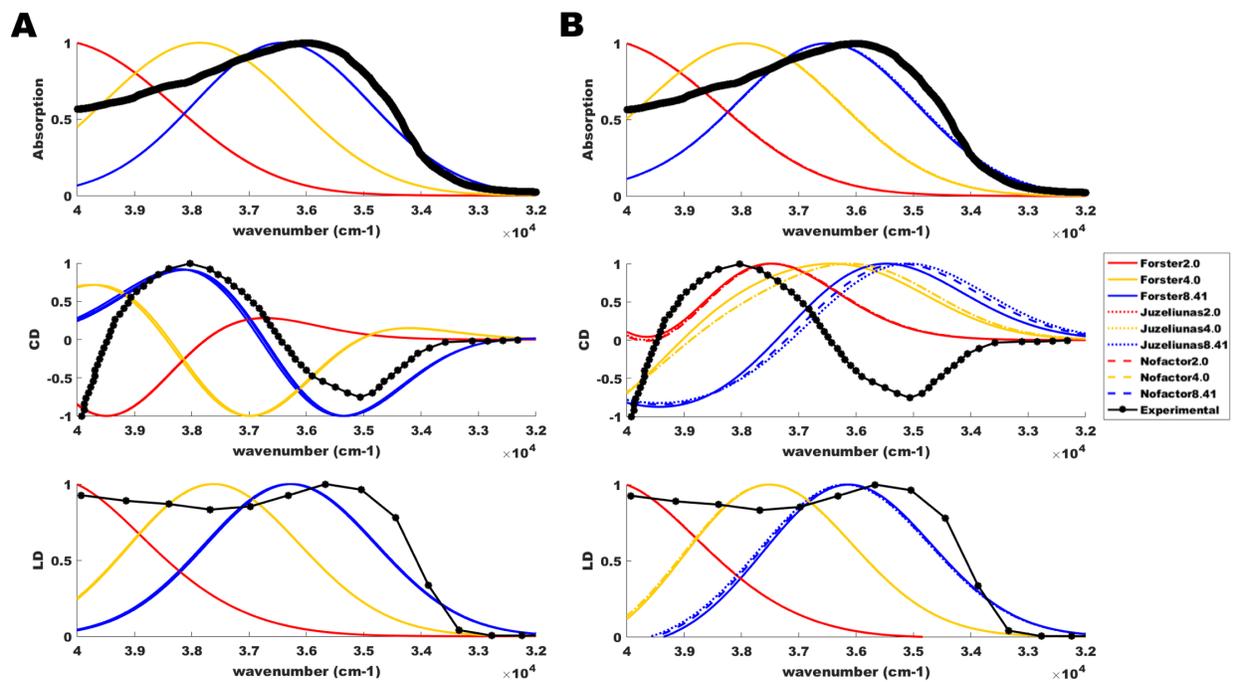

Figure 5

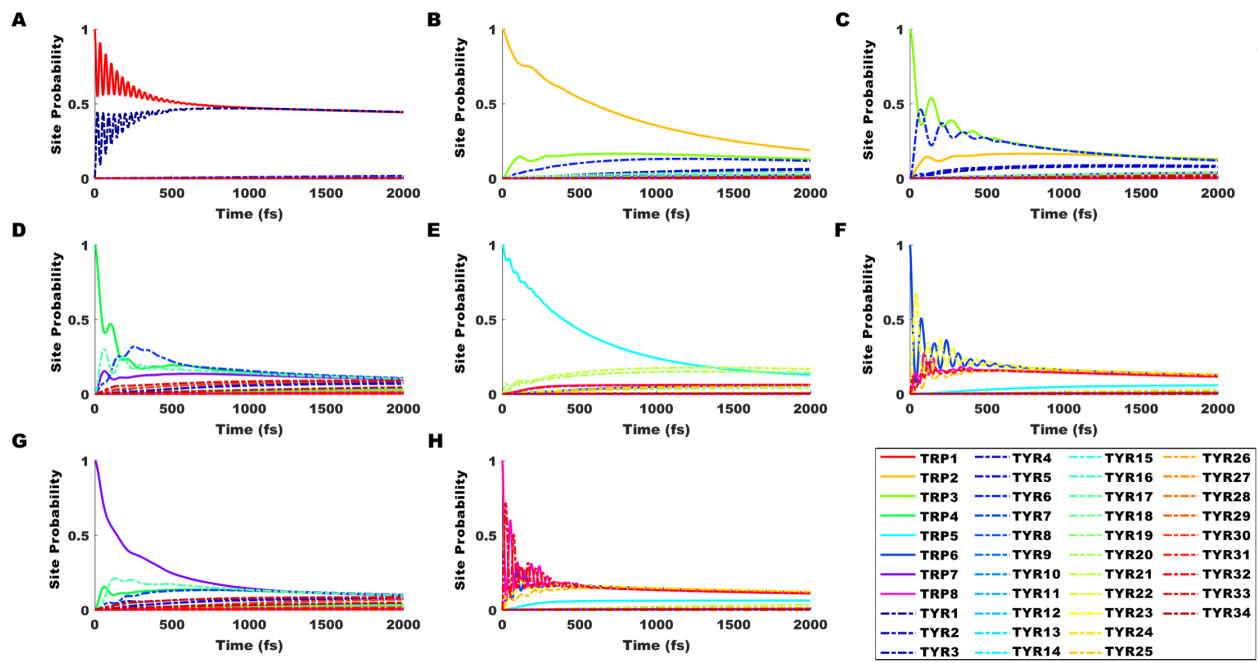

Figure 6

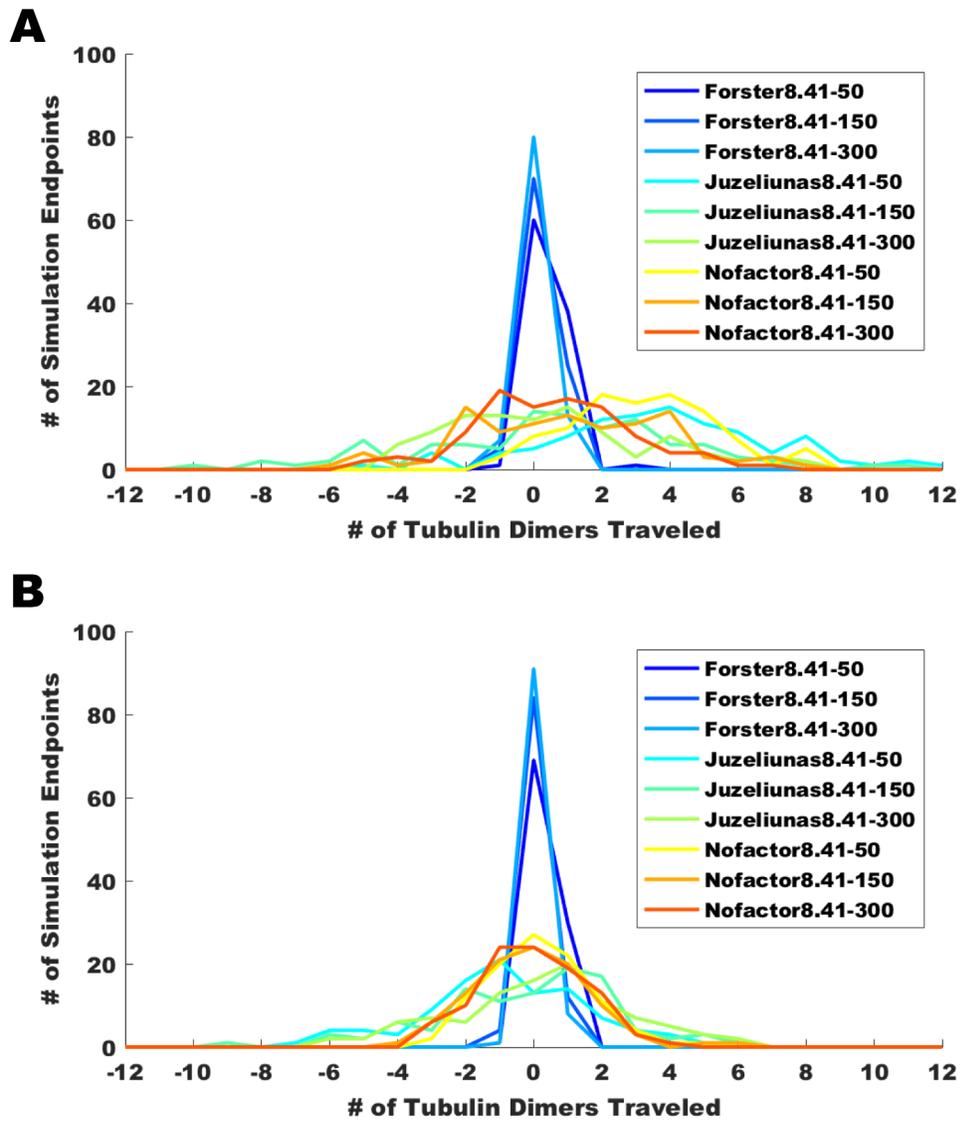

Figure 7

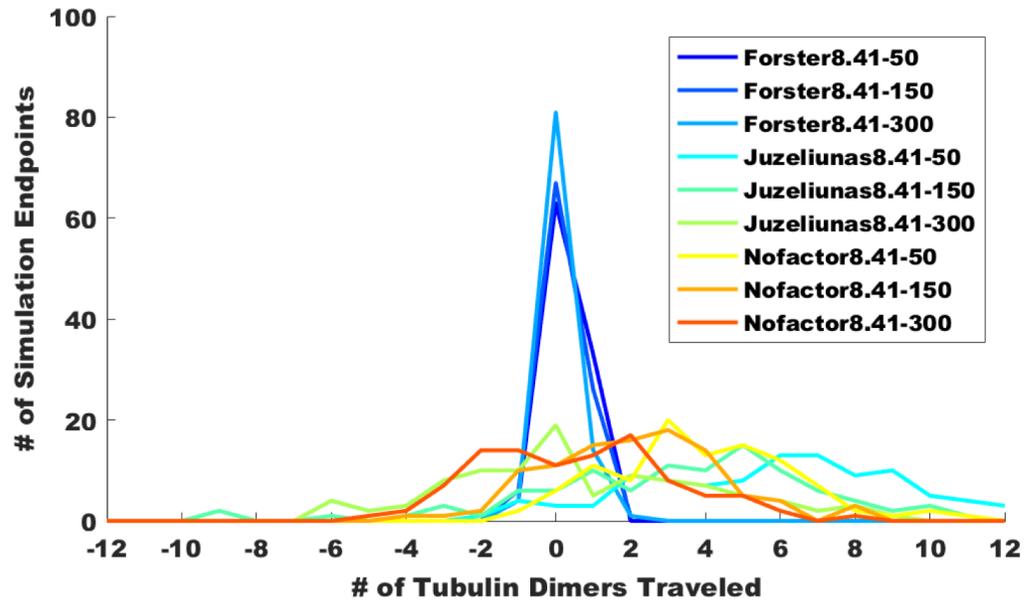